# Oxygen-Pressure-Limited Recovery of the Hematite α-$Fe_2O_3$(0001) Surface from a Reduced $Fe_3O_4$(111)-Like Layer

*Nishant Kumar,[1] Matthias Blatnik,[1] Jan Čechal[1,2]**

[1] CEITEC - Central European Institute of Technology, Brno University of Technology, Purkyňova 123, 612 00 Brno, Czech Republic.

[2] Institute of Physical Engineering, Brno University of Technology, Technická 2896/2, 616 69 Brno, Czech Republic.

## Abstract

The oxidation kinetics of hematite α-$Fe_2O_3$(0001) surfaces are vital for its applications in catalysis, environmental remediation, and industrial processes. Despite prior studies, the roles of temperature, oxygen partial pressure, and oxygen chemical potential in controlling nucleation and growth kinetics are not fully understood. Using real-time Low Energy Electron Microscopy/Diffraction (LEEM/LEED), we systematically investigate the oxidation of a reduced $Fe_3O_4$(111)-like surface layer to hematite under controlled conditions. We show that complete recovery of the hematite surface termination is closely linked to the nucleation and lateral growth of a two-dimensional honeycomb (H) phase. While higher temperatures accelerate nucleation, they slow lateral growth at constant oxygen pressure, indicating that oxygen supply limits the oxidation rate. Below an oxygen partial pressure threshold ($\sim 2\times10^{-6}$ mbar), growth dramatically slows, underscoring the critical role of oxygen availability. Below a certain oxygen pressure threshold, the growth time rapidly increases. Our study elucidates the interplay between thermodynamics and kinetics in hematite surface oxidation, informing strategies to optimize surface properties for catalytic and industrial processes.

**Introduction**

Hematite ($\alpha$-$Fe_2O_3$) is an abundant, low-toxic, catalytic, and photocatalytic material employed for CO oxidation, water purification, liquid fuel synthesis, styrene production, and water splitting.[1–3] Because catalytic reactions occur on the surface of materials, a detailed understanding of surface structure and properties is essential for a reliable description and modeling of surface reactions. Accordingly, the hematite $\alpha$-$Fe_2O_3$(0001) surface has been extensively studied using surface science techniques under ultrahigh vacuum (UHV) environment.[1] While UHV conditions enable atomic-level insights, these studies are necessarily limited to low pressures. When the sample is treated at a low oxygen chemical potential below $\mu_O < -2.5$ eV or subjected to oxygen-selective sputtering during surface preparation, hematite $\alpha$-$Fe_2O_3$ is reduced,[1,4,5] resulting in the formation of an $Fe_3O_4$(111) layer that is epitaxially matched to the underlaying hematite $\alpha$-$Fe_2O_3$(0001). This process creates an atomically sharp $Fe_3O_4$(111)/$\alpha$-$Fe_2O_3$(0001) interface.[6] Subsequent annealing under appropriate $O_2$ partial pressures at $\mu_O > -2.5$ eV restores stoichiometric $\alpha$-$Fe_2O_3$(0001),[1] which is terminated by a two-dimensional layer referred to as the honeycomb phase (H-Phase),[6,7] a reconstructed overlayer with a characteristic moiré pattern. Within this chemical potential range, multiple bulk and surface phases coexist, complicating the preparation of well-defined samples.[1] In addition to thermodynamic factors, the oxidation temperature and oxygen partial pressure strongly influence the oxidation kinetics; full oxidation of a reduced surface layer can require from tens of minutes to several hours, depending on conditions.[5,6,8] Despite its importance for preparing well-defined surfaces, the kinetics of this transformation have not been quantitatively studied.

In this work, we employ low-energy electron microscopy (LEEM)[9–14] to monitor oxidation $Fe_3O_4$(111)-like layers on $\alpha$-$Fe_2O_3$(0001) in real space and time. We systematically investigate how nucleation and growth of the H-phase depend on oxygen partial pressure and temperature, revealing that the kinetics do not follow simple Arrhenius (temperature-activated) . Notably, we find that oxidation becomes slower with increasing temperature when the oxygen partial pressure is held constant, highlighting the critical role of oxygen supply.

Both hematite and magnetite are iron oxides with close-packed $O^{2-}$ anion sublattices, but differ in their Fe cation arrangements.[1] Hematite α-$Fe_2O_3$ has a corundum structure with $Fe^{3+}$ in octahedral sites. Magnetite $Fe_3O_4$ has an inverse spinel structure where $Fe^{3+}$ ions occupy tetrahedral sites, and both $Fe^{2+}$ and $Fe^{3+}$ ions share octahedral sites in a 1:1 ratio. Upon oxidation, $Fe^{2+}$ is converted to $Fe^{3+}$ within the spinel structure, leading to the formation of Fe vacancies in the octahedral sublattice; if the spinel framework is preserved during this process, the resulting phase is maghemite γ-$Fe_2O_3$ with a cation-deficient spinel structure.[1]

Magnetite to hematite transformation, known as martitization, is significant in geological sciences[15] and magnetite ore processing.[15,16] Two main martitization pathways are recognized: (i) an oxidizing reaction requiring molecular oxygen and (ii) a non-redox acid-base hydrothermal replacement reaction in contact with a liquid water environment via aqueous ferric species.[17,18] In the oxidizing pathway, martitization proceeds in two-steps: formation of maghemite γ-$Fe_2O_3$ via Fe cation vacancy diffusion[19–22] followed by the gradual transformation of maghemite γ-$Fe_2O_3$ to hematite α-$Fe_2O_3$.[1,17,23] The conversion of metastable γ-$Fe_2O_3$ into α-$Fe_2O_3$ involves a structural rearrangement from the face-centered cubic (fcc) spinel $O^{2-}$ sublattice to the hexagonal close-packed (hcp) corundum anion sublattice.

Further insights into magnetite oxidation have been obtained from LEEM studies on the (100) surface termination.[12,13] Under low oxygen partial pressure (~$10^{-6}$ mbar) and temperatures around 650 °C, Fe cation diffusion is mediated by Fe vacancies.[21,22] Magnetite to hematite transformation proceeds through two spatially separated reactions:[12] at the magnetite surface $O_2$ molecules dissociate and oxidize $Fe^{2+}$ to $Fe^{3+}$, which leads to the formation of iron vacancies in magnetite. These vacancies readily diffuse over micrometer distances within the magnetite lattice[12,19] and subsequently facilitate the formation of hematite via a topotactic reaction that conserves oxygen. The ionic flow is balanced by a flow of electrons enabled by the good electrical conductivity of magnetite.[24] In the case of $Fe_3O_4(100)$ oxidation, the hematite forms inclusions inclined by 45° with respect to the surface plane, since the $Fe_3O_4\{111\}$ planes match the α-$Fe_2O_3(0001)$ plane, coinciding with the sharp interface between the $Fe_3O_4(111)$ layer on α-$Fe_2O_3(0001)$ bulk observed by TEM.[6] The role of intermediate maghemite formation remains unclear. While there was no evidence for intermediate maghemite formation from electron spectroscopies,[13] studies of 5-nm $Fe_3O_4(111)$ films on Pt(111) have shown that the films initially become non-stoichiometric toward maghemite, and even after hematite forms at the surface, the underlying layers retain the spinel structure.[25] Under oxidizing conditions, the equilibrium concentration of cation vacancies lowers with increasing temperature.[22,26] At high temperatures (900 °C – 1400 °C), this concentration increases with pressure $p_{O_2}$ as $\left(\frac{p_{O_2}}{p^*}\right)^{2/3}$, with $p^*$ being atmospheric pressure.[22] At lower temperatures (200 °C – 400 °C), the Fe vacancy concentration is much lower than bulk equilibrium values.[19] For $Fe_3O_4(100)$ oxidation, the reaction rate significantly decreased for oxygen partial pressures below $1.3\times10^{-6}$ mbar,[12] likely due to the kinetics of vacancy formation tather than to their equilibrium concentration. Because Fe vacancies diffuse readily, the oxidation rate of a 20 nm $Fe_3O_4(111)$ layer on α-$Fe_2O_3(0001)$ bulk we do not expected to dependend significantly on layer thickness in contrast with oxidation of metal layers as described by the classical Cabrera-Mott theory.[27]

By employing real-time LEEM, we were able to monitor surface transformations with high spatial and temporal resolution, capturing both the nucleation and growth of the two-dimensional honeycomb (H-phase) surface layer during oxidation. In this study, we systematically investigate how oxidation kinetics are affected by independently controlling partial oxygen pressure ($p_{O_2}$), temperature, and oxygen chemical potential. This approach allows us to disentangle the individual contributions of these parameters to the nucleation and growth processes, thereby clarifying their roles in the overall reaction kinetics.

## Results

### Initial Surface Preparation and Characterization

To establish a stable and controlled starting surface for the oxidation studies, we used an α-$Fe_2O_3$(0001) hematite single crystal featuring the 2D honeycomb phase fully covering the surface. This surface is referred to as the H-phase in the following text. Samples featuring H-phase were subsequently reduced to magnetite $Fe_3O_4$(111) through four cycles of $Ar^+$ sputtering and UHV annealing at 640°C (see Methods section for details). Sputtering selectively removes surface oxygen atoms, which reduces the near-surface volume to stoichiometric $Fe_3O_4$, forming a ~20-25 nm thick layer that maintains the epitaxial relationship with the bulk $Fe_2O_3$ hematite.[6] We will refer to this surface as the R-phase in the following text.

The transition between the H- and R-phases was characterized using X-ray Photoelectron Spectroscopy (XPS) and Low Energy Electron Microscopy/Diffraction (LEEM/LEED). In the XPS spectrum of the R-phase (Figure 1a), the $Fe^{3+}$ satellite peaks are absent, whereas the spectrum of the H-phase (Figure 1b), these satellite peaks are present, confirming the oxidation state.[28–30] LEEM/LEED provide complementary real and reciprocal-space views, elucidating the sample morphology at the mesoscale and its structure, respectively. Bright-field LEEM imaging revealed a uniform and well-prepared R-phase surface (Figure 1c).[7] The corresponding diffraction pattern in Figure 1d displays (2×2)-$Fe_3O_4$(111) diffraction spots (referred to the O planes[6]) characteristic of this surface[6,7,31] , characteristic of $Fe_3O_4$(111) terminated at the tetrahedral Fe plane.[1] The oxidation of the surface at temperatures of 660 – 700°C and oxygen partial pressures of $6.3\times10^{-7}$ – $5.0\times10^{-6}$ mbar leads to full oxidation and to α-$Fe_2O_3$(0001) featuring the H-phase; depending on the oxidation conditions, it takes 55 to 450 minutes. The bright-field LEEM image of the H-phase (Figure 1e) shows a uniform surface, and the diffraction pattern (Figure 1f) exhibits the characteristic flowered pattern[29,31,32] around the bulk spots associated with the moiré due to the presence of the overlayer.[7] We note that the maghemite surface can exhibit the same (2×2) surface reconstruction as magnetite;[33] therefore, maghemite and magnetite cannot be distinguished by simple diffraction experiments alone.

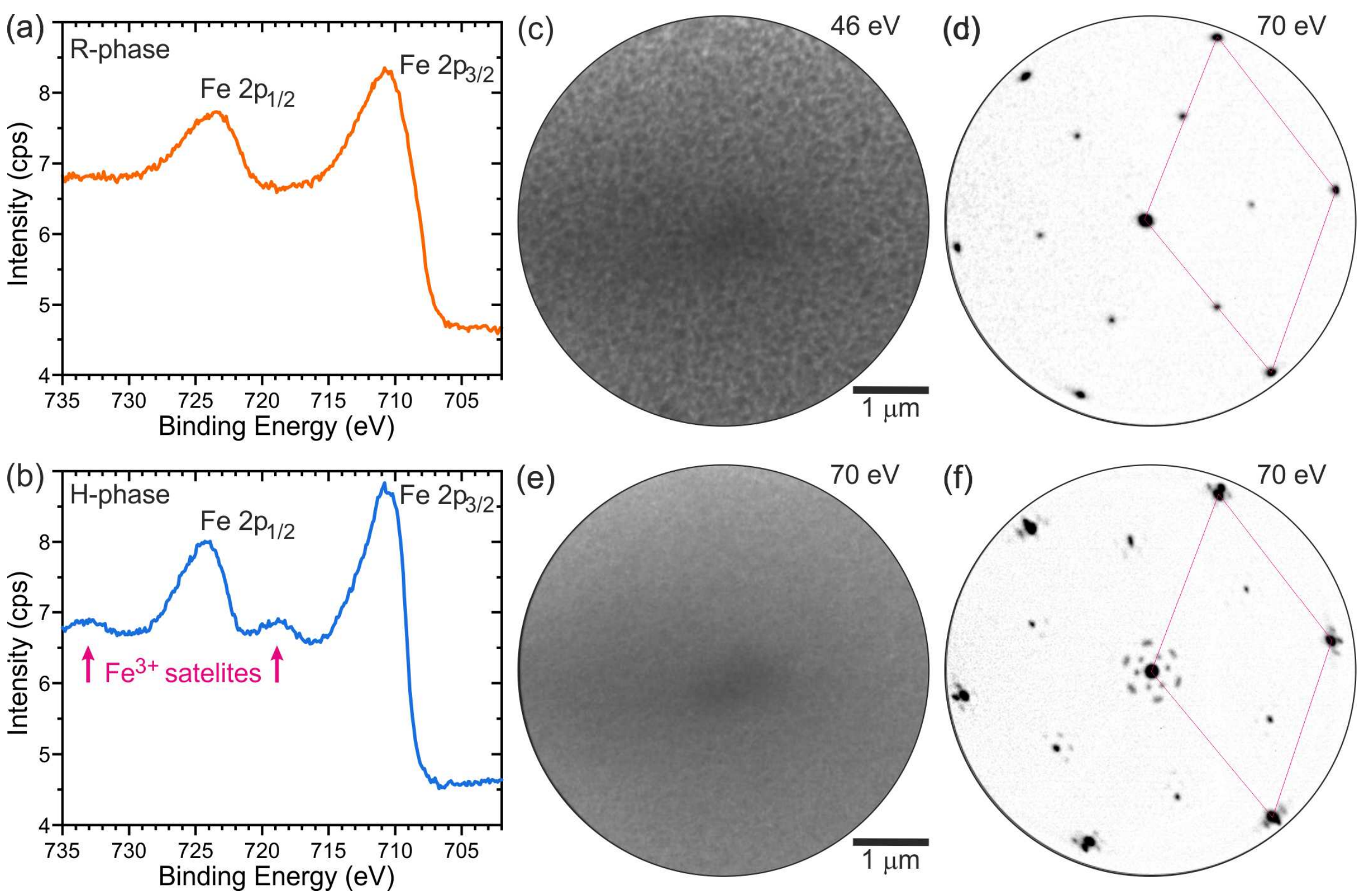

**Figure 1:** Surface characterization of hematite (H-phase) and reduced magnetite (R-phase). (a) XPS spectrum of the R-phase, confirming the absence of $Fe^{3+}$ satellite peaks and the successful reduction of α-$Fe_2O_3$ to $Fe_3O_4$. (b) XPS spectrum of the H-phase, highlighting the presence of $Fe^{3+}$ satellite peaks characteristic of hematite. (c) Bright-field LEEM image of the magnetite (R-phase) surface, showing a uniform and well-prepared morphology. (d) (2×2) diffraction pattern at 70 eV characteristic for the R-phase, indicative of the well-ordered surface suitable for oxidation studies. (e) Bright-field LEEM image at 70 eV showing the H-phase surface. (f) Diffraction pattern at 70 eV characteristic of the H-phase. The (1×1) unit cell is marked by a rhombus in (d) and (f).

**Time Evolution of Oxidation**

To obtain quantitative insights into the R- to H-phase transformation, we performed three sets of real-time LEEM/LEED experiments, keeping constant either partial oxygen pressure $p_{O_2}$, sample temperature $T$, or oxygen chemical potential $\mu$, while varying either temperature or pressure, or in the last case, both parameters. By systematically separating the effects of these parameters, we revealed their impact on the R- to H-phase oxidation kinetics. An example of the oxidation kinetics at 660°C and $6.3\times10^{-7}$ mbar ($\mu_O = -1.867$ eV) using real-time LEEM/LEED is given in Figure 2. During the first stage of oxidation, the entire surface shows a uniform contrast and the diffraction pattern of the R-phase. At 75 min, moiré spots became visible, which indicates H-phase nucleation (Figure 2a). We refer to this time, $t_0$, as the nucleation time in the following. As oxidation progresses, the H-phase spots become more pronounced, while the R-phase spots progressively fade.

The bright-field LEEM image in Figure 2b illustrates the progressive transformation of the surface from the reduced R-phase to the H-phase. After nucleation, H-phase islands emerge and remain surrounded by the R-phase. As the oxidation advances, the R-phase regions on the surface gradually decrease, while the H-phase regions expand. By 355 min, the entire surface is covered by the H-phase, marking the completion of the oxidation process. Quantification of the evolution of H-phase coverage derived from real-space LEEM images over time is given in Figure 3. In this figure, we also define all characteristic times. Nucleation time $t_0$ is marked by the green dashed line. Time $t_{100}$, defined as the time from the start of oxidation until complete transformation, is referred to as the total time to full oxidation; we will use the derived parameter $t_{100} - t_0$, giving the H-phase growth time. These parameters are shown in Figure 4 for experiments in which the pressure, temperature, or chemical potential was kept constant. The complementary reciprocal quantities, i.e., inverse nucleation time and growth rate, are provided in Figure S1 in Supporting Information. A supplementary parameter, the characteristic growth time $t_{50}$, was obtained as the inverse of the growth rate determined from the slope of a linear fit of the coverage–time dependence around 50 % H-phase surface coverage. The characteristic growth time yields results consistent with the H-phase growth time; it is shown in Figure S2 in Supporting Information. All the values are summarized in Table S1.

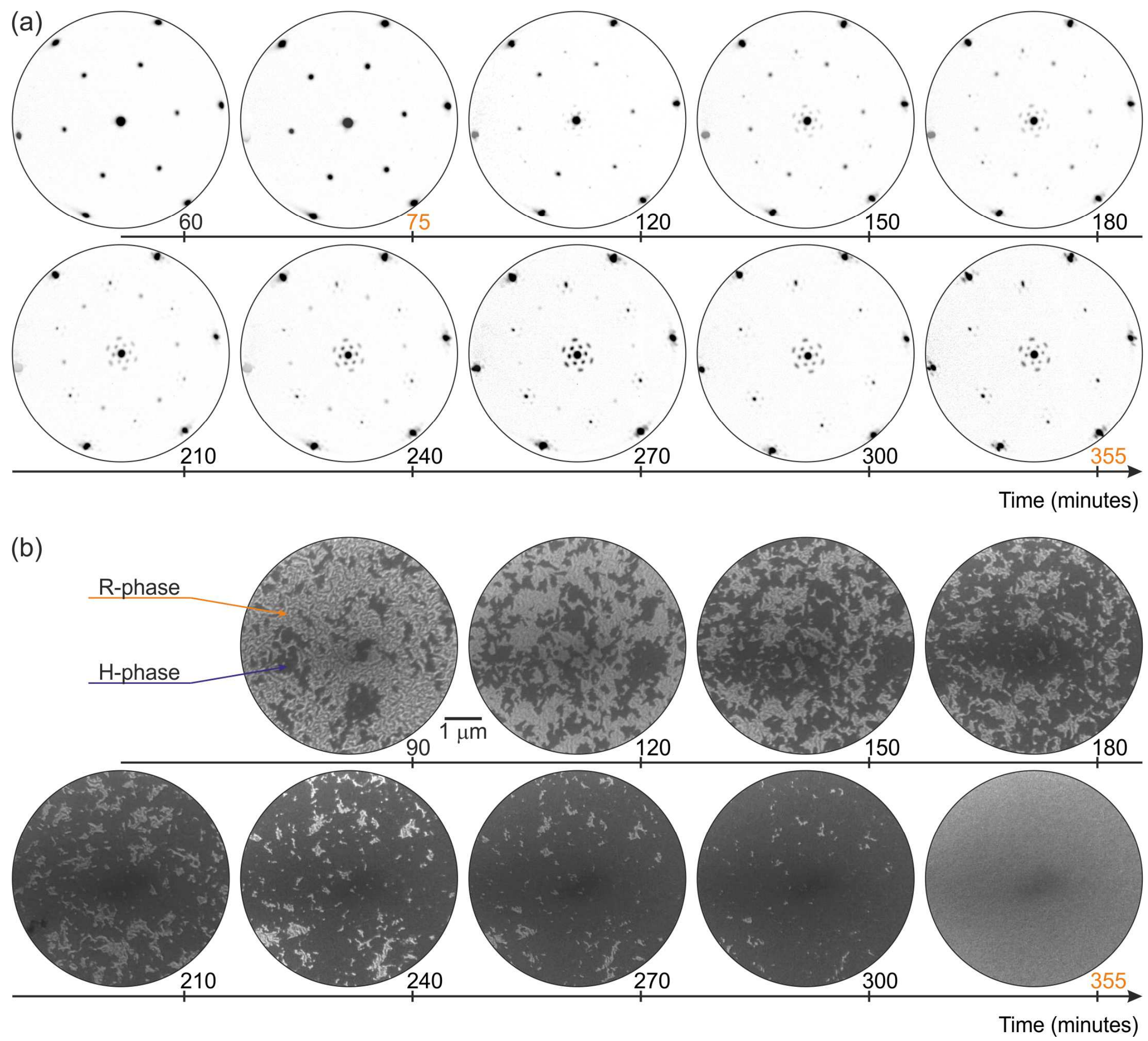

**Figure 2:** Surface evolution during the oxidation from magnetite to hematite at 660°C and $6.3\times10^{-7}$ mbar in real and reciprocal space. (a) Sequence of diffraction patterns taken at 70 eV showing the structural transformation from the R-phase to the H-phase. (b) Corresponding sequence of bright-field LEEM images capturing the transformation from the R-phase (brighter) to the oxidized H-phase (darker) over 355 minutes.

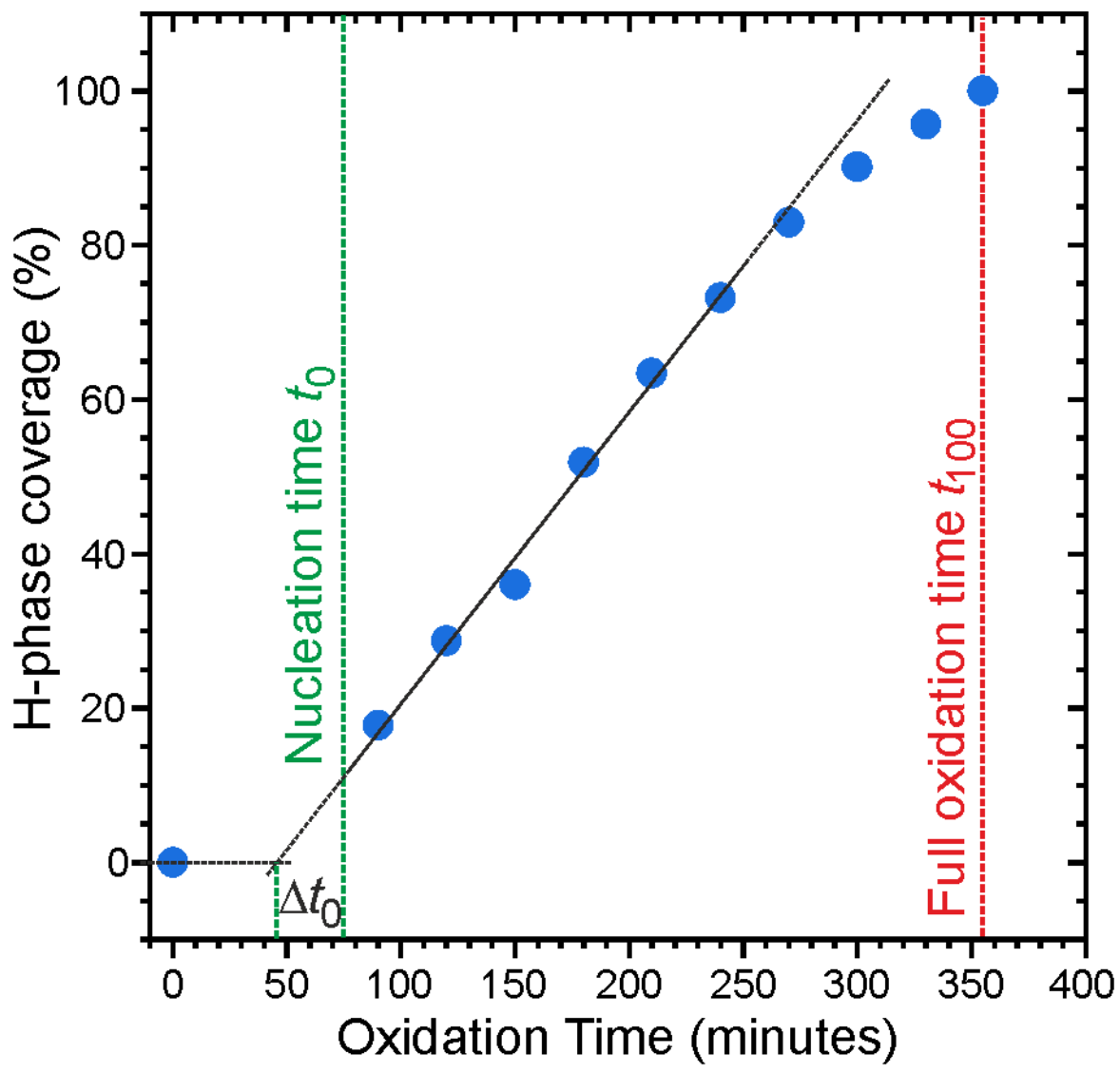

**Figure 3**: Coverage vs. time graph quantifying the H-phase surface coverage (%) from the sequence of the bright-field LEEM images. during oxidation from magnetite to hematite at 660°C and 6.3×10$^{-7}$ mbar ($\mu_O = -1.867$ eV). The marked nucleation time is the first occurrence of the H-phase-related diffraction spots. The full oxidation marks the full substrate covered in the bright-field LEEM. The error in nucleation time was defined as the difference between t0 and the time extrapolated from the linear increase to zero coverage.

**Oxidation at Constant Oxygen Partial Pressure**

We applied the real-time real-space LEEM/LEED methodology described above to study the oxidation kinetics under a range of temperatures and pressures. We start with the standard experiment in which we have varied the temperature and kept the oxygen partial pressure constant ($p_{O_2} = 1.8\times10^{-6}$ mbar). The characteristic time $t_0$ for the onset of nucleation, and the total growth time $t_{100}$- $t_0$, as a function of temperature, are displayed in Figure 4a,b.

Nucleation time $t_0$ slightly decreased with increasing temperature. At 660°C, nucleation required 35 min, whereas at 700°C, it decreased to 26 min. This behavior aligns with thermally activated processes, in which higher temperatures enhance the rate of the limiting process.

Conversely, the growth time $t_{100}$-$t_0$ increased with temperature. The total growth time $t_{100}$-$t_0$ increases approximately linearly with temperature; at 660°C, the H-phase growth time was 55 min, whereas it extended to 94 min at 700°C (Figure 4b). The associated reciprocal times show approximately linear increases in the inverse nucleation time and a decrease in the growth rate with temperature (Figure S1 in Supporting Information).

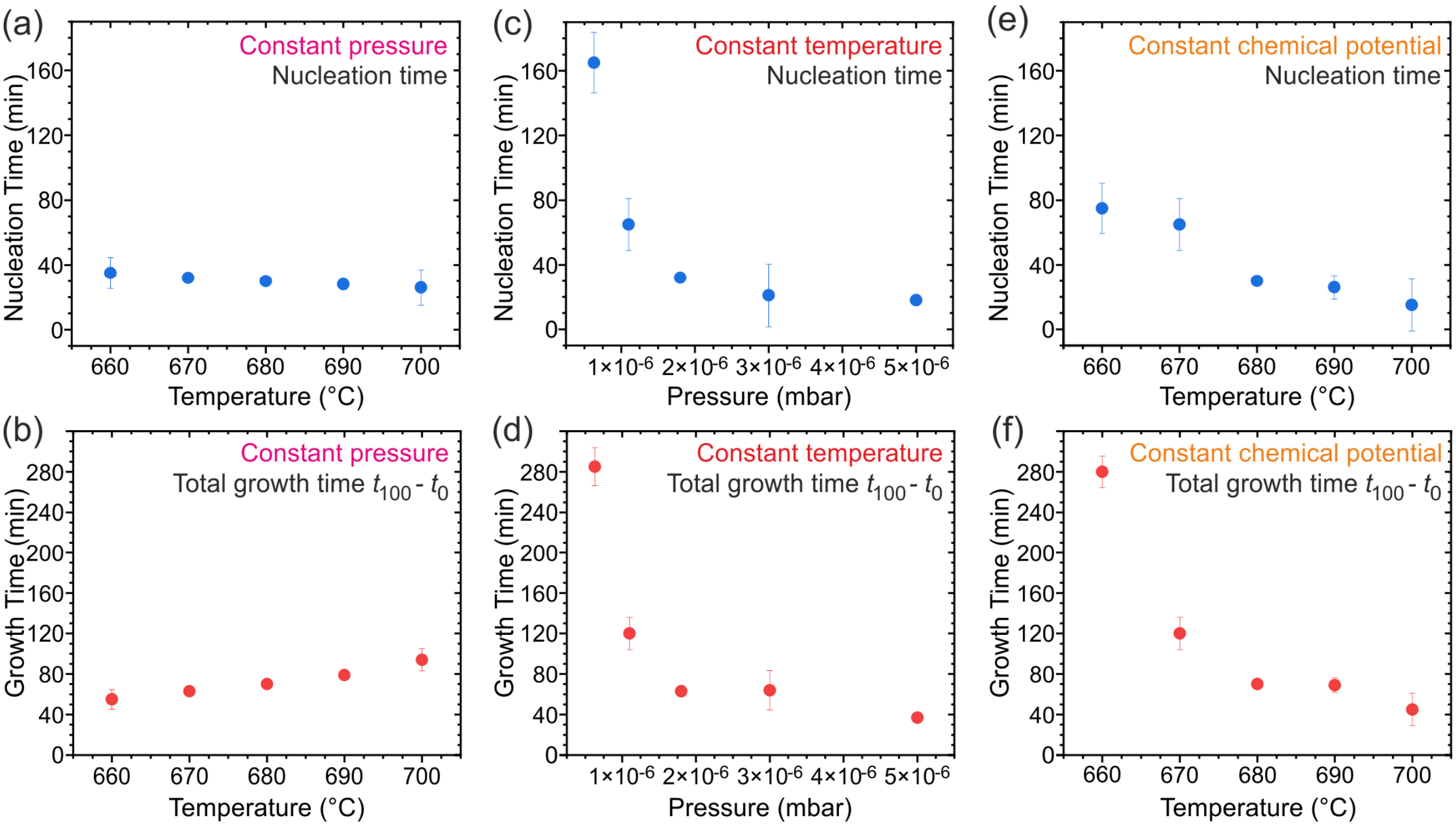

**Figure 4:** Characteristic R-phase to H-phase transformation times under varying experimental conditions. (a, b) Nucleation time ($t_0$) and total growth time ($t_{100}$- $t_0$) as functions of temperature at constant $p_{O_2} = 1.8 \times 10^{-6}$ mbar. (c, d) Nucleation time ($t_0$) and total growth time ($t_{100}$- $t_0$) as functions of $p_{O_2}$ at a constant temperature of 670°C. (e, f) Nucleation time ($t_0$) and total growth time ($t_{100}$- $t_0$) as functions of temperature under constant oxygen chemical potential $\mu_O = -1.867$ eV. Error bars represent one standard deviation, reflecting the uncertainty of deriving the given parameter, if the error bars are small compared to the symbol size, they are not displayed.

## Oxidation at Constant Temperature

We observe that the increasing temperature alone does not result in an increased oxidation rate. To further investigate this, we examined the oxidation kinetics under a constant temperature of 670°C while systematically varying the $p_{O_2}$. Changes in oxygen pressure influence both nucleation and growth. Both nucleation time $t_0$ and growth time $t_{100}$ - $t_0$ decreased with increasing $p_{O_2}$ as shown in Figures 4c,d. At the lowest pressure of 6.3×10$^{-7}$ mbar, nucleation required approximately 165 min, decreasing significantly to 18 min at the highest pressure of 5.0×10$^{-6}$ mbar. Similarly, the H-phase growth time $t_{100}$ - $t_0$ (Figure 4d) decreases from 285 to 37 minutes with increasing pressure. There is a dramatic increase in both times in the low-pressure range, most clearly below ~2.0×10$^{-6}$ mbar. Considering the reciprocal times (the inverse time for nucleation and the growth rate), these increase with pressure in the low-pressure regime and tend to saturate at higher pressures (Figure S1 in Supporting Information). Thus, low oxygen pressure significantly limits the oxidation rate, while further increases above a certain threshold have a diminishing effect.

## Oxidation at Constant Oxygen Chemical Potential

Informed by the fact that oxygen availability is a critical factor, we performed a temperature-dependent study, but under a constant oxygen chemical potential, by simultaneously varying pressure to keep the oxygen chemical potential constant at $\mu_O = -1.867$ eV according to[34] $\mu_O = 1/2\left[\mu^0_{O_2}(T) + kT\, ln\left(p_{O_2}/p^0\right)\right]$, where $k$ is the Boltzmann constant, and $T$ is the absolute temperature in Kelvin, $p_{O_2}$ is the oxygen partial pressure, and $p^0 = 1$ bar. The reference chemical potential $\mu^0_{O_2}(T) = H_{O_2}(T,p^0) - H_{O_2}(0,p^0) - T\left[S_{O_2}(T,p^0) -\right.$

$S_{O_2}(0,p^0)]$ is chosen so that $\mu_{O_2}(0,p_{O_2}) = 0$;[35] the enthalpy $H_{O_2}(T,p^0)$ and entropy $S_{O_2}(T,p^0)$ per $O_2$ molecule are taken from thermochemical tables.[36]

At constant chemical potential, the nucleation times $t_0$ exhibited a stronger temperature dependence than at constant $p_{O_2}$. It decreased from 75 min to 15 min when the temperature increased from 660°C to 700°C (Figure 4e). The constant-pressure study was performed at an intermediate pressure of $1.8\times10^{-6}$ mbar with the chosen chemical potential corresponding to this pressure at 680 °C. To keep the chemical potential constant, lowering the temperature must be accompanied by a decrease in pressure, which prolongs nucleation time. Conversely, at higher temperatures, the pressure is higher, and the nucleation time is decreased. The H-phase growth time $t_{100}$ - $t_0$ (Figure 4f) first decreases from 280 to 70 minutes, followed by a slower decrease to 45 minutes beyond 680 °C. These findings emphasize the strong role of oxygen partial pressure and oxygen supply, even when the thermodynamic driving force is kept constant.

**Discussion**

The initial state is a layer of magnetite on bulk hematite. Hence, during the oxidation, the reduced layer should be re-oxidized, and the overlayer termination formed (H-Phase). BF-LEEM imaging reveals that H-Phase islands nucleate at multiple locations and grow laterally. Throughout oxidation, the H-phase islands remain surrounded by R-phase, which only disappears once the surface is fully covered by H-phase. The [0001] direction of hematite exhibits a polar alternation of Fe and O planes, necessitating compensation of the resulting surface polarity. This compensation can occur via impurity segregation or by the H-phase termination.[6] Across all experimental conditions, we consistently observe that complete recovery of the hematite surface termination is closely associated with the presence of the H-phase layer.

Our data also show that the growth does not follow simple thermally activated (Arrhenius) kinetics, as the supply of oxygen for the reaction is a critical factor. When the temperature is increased at constant oxygen pressure, the nucleation rate increases but the growth rate decreases, indicating that nucleation and growth are governed by distinct kinetic pathways. The observed decrease in the growth rate suggests that the reaction is not solely controlled by thermal activation, but is also limited by the oxygen availability. To enhance the growth rate, the increase in temperature should be accompanied by a corresponding increase in oxygen partial pressure. On the other hand, the growth rate decreases dramatically at oxygen partial pressures below $2\times10^{-6}$ mbar. In this pressure range, the reciprocal growth time increases with pressure, while further increases have a diminishing effect on the kinetics. Under these low-pressure conditions, maintaining constant chemical potential by lowering the temperature does not compensate for the limiting effect of low pressure, as the pressure dependence dominates. A similar decrease for oxygen partial pressures below $1.3\times10^{-6}$ mbar was also observed in the case of $Fe_3O_4$(100) oxidation.[12]

Considering the oxidation mechanism, the key step is the formation of Fe vacancies via oxygen dissociation at the magnetite surface, followed by their subsequent transport.[12] Previous studies on the $Fe_3O_4$(100) have shown that the oxidation rate is limited by the kinetics of Fe vacancy formation rather than their equilibrium concentration. The observed increase in growth time with increasing temperature thus suggests a competing process, which could be either $O_2$ desorption during oxygen splitting or the diffusion of Fe vacancies into the bulk. As shown in Figure 3, the growth rate reamains uniform until the final stages of oxidation. Given that dissociative oxygen adsorption occurs uniformly across the magnetite surface,[12] significant local variations in growth rate would be expected if the supply of dissociated oxygen from terraces were the sole limiting factor. This observation indicates that the transport of reaction-mediating defects, such as Fe vacancies, to the reaction site is a critical factor in determining the overall oxidation rate.

Our results suggest a more complex nature of the growth of the H-phase in which two distinct oxidations are involved, that is, oxidation of bulk magnetite and formation of the terminating layer of H-phase, with bulk oxidation not appearing to be completed at the surface in the absence of the H-phase. Elucidating the relationship between these two oxidation processes would benefit from a combined approach using LEEM and cross-sectional TEM imaging,[6] which we plan to pursue in future work.

**Conclusions**

In conclusion, real-time LEEM/LEED shows that oxidation of the reduced $Fe_3O_4(111)$-like surface layer on α-$Fe_2O_3(0001)$ proceeds through nucleation and lateral growth of the H-phase. The oxidation kinetics show a significant dependence on oxygen supply for the surface reaction. While a slight increase in temperature slightly accelerates H-phase nucleation, lateral growth slows at constant oxygen pressure, indicating that the supply of reactive oxygen limits the surface reaction. Increasing temperature must therefore be accompanied by a sufficient increase in oxygen partial pressure to sustain the oxidation rate. Below an oxygen partial pressure of $\sim 2\times10^{-6}$ mbar, the H-phase growth rate decreases dramatically, and this decrease cannot be compensated simply by lowering the temperature. The observation that residual R-phase persists until H-phase coalescence shows that complete recovery of the hematite surface termination is closely coupled to lateral H-phase growth.

## Methods

Sample preparation, LEEM, and XPS characterization were performed in an ultrahigh-vacuum (UHV) system at the CEITEC Nano Research Infrastructure. The system comprises multiple UHV chambers interconnected by a UHV transfer line, allowing sample transfer between chambers via the UHV transfer line, maintaining a base pressure of $2\times10^{-10}$ mbar.

**Sample Preparation.**

A natural $\alpha$-$Fe_2O_3(0001)$ single crystal was acquired from SurfaceNet GmbH. To avoid potential interference from impurities in natural crystals, synthetic hematite films were prepared by pulsed laser deposition (PLD) in a dedicated setup with *in situ* ultra-high vacuum (UHV) transfer to a surface-science chamber. The film of ~100 nm thickness was grown on the natural $\alpha$-$Fe_2O_3(0001)$ single crystal from an iron oxide target using a KrF excimer laser (Coherent Compex Pro 201, $\lambda = 248$ nm) at 5 Hz pulse frequency and 2.0 J $cm^{-2}$ laser fluence. The substrate was heated from the back by a 980 nm continuous-wave infrared laser (DILAS Compact Evolution), and its temperature was monitored with an Impac IGA5 pyrometer. Deposition was performed at 850°C substrate temperature in $2\times10^{-2}$ mbar $O_2$, with a heating ramp rate of 35°C $min^{-1}$.

Prior to film deposition, the natural sample underwent cleaning by multiple sputtering-annealing cycles: 10 min of sputtering with $5\times10^{-7}$ mbar $Ar^+$ at 1 keV, followed by 10 min of annealing at 600°C in $1\times10^{-6}$ mbar $O_2$. This procedure was repeated until no further changes in contaminant signals were observed in the XPS analysis. Subsequently, the sample was annealed at 800°C for 1 hour under $5\times10^{-5}$ mbar $O_2$ to facilitate surface morphology flattening. The cleanliness was confirmed by LEED and XPS.

The sample featuring the synthetic film was used for all experiments. To reduce the as-prepared $\alpha$-$Fe_2O_3$(0001) near-surface volume to the magnetite phase ($Fe_3O_4$), the sample was subjected to four cycles of $Ar^+$ sputtering (10 min cycle, $9\times10^{-6}$ mbar, 1keV, 10 mA) followed by annealing in UHV at 640°C for 10 min.

**Low-Energy Electron Microscopy/Diffraction (LEEM/LEED)** experiments were conducted using a Specs FE-LEEM P90 instrument. Bright-field images were measured by capturing electrons collected from the central (0,0) diffracted beam. Diffraction patterns were acquired from a surface area of approximately $7 \times 15$ $\mu m^2$ without aperture confinement. The coverage of the H-phase was determined using ImageJ software as the difference between the areas of the R- and H-phase regions.

**X-ray photoelectron Spectroscopy (XPS)** analysis was performed using a Specs system equipped with a Phoibos 150 spectrometer. Measurements were performed with a non-monochromatized X-ray source emitting Mg Kα radiation (hν = 1253.6 eV), at 0° emission angle, utilizing the high-magnification mode with a 15 mm iris aperture. Core-level spectra were acquired with a pass energy of 20 eV, a step size of 0.1 eV, and a dwell time of 0.1 s. No charging effects were observed during the measurements.

## ASSOCIATED CONTENT

Supplemental Material: a table summarizing the experimental data and figures showing the nucleation/oxidation rates and dependence of parameter $t_{50}$.

## AUTHOR INFORMATION

**Corresponding Author**

* E-mail: cechal@fme.vutbr.cz (J. Č.)

## COMPETING INTERESTS

The authors declare no competing interests.

## ACKNOWLEDGMENT

M.B. was supported by the ESF under the project CZ.02.01.01/00/22_010/0002552 and N.K. by the Specific Research project of Brno University of Technology. We acknowledge CzechNanoLab Research Infrastructure (LM2023051) supported by MEYS CR.

## DATA AVAILABILITY

The data that support the findings of this study are available from the corresponding author upon reasonable request.

REFERENCES

(1) Parkinson, G. S. Iron Oxide Surfaces. *Surf. Sci. Rep.* **2016**, *71* (1), 272–365. https://doi.org/10.1016/j.surfrep.2016.02.001.

(2) Sivula, K.; Le Formal, F.; Grätzel, M. Solar Water Splitting: Progress Using Hematite (A-Fe 2 O 3 ) Photoelectrodes. *ChemSusChem* **2011**, *4* (4), 432–449. https://doi.org/10.1002/cssc.201000416.

(3) Tamirat, A. G.; Rick, J.; Dubale, A. A.; Su, W.-N.; Hwang, B.-J. Using Hematite for Photoelectrochemical Water Splitting: A Review of Current Progress and Challenges. *Nanoscale Horizons* **2016**, *1* (4), 243–267. https://doi.org/10.1039/C5NH00098J.

(4) Tang, Y.; Qin, H.; Wu, K.; Guo, Q.; Guo, J. The Reduction and Oxidation of Fe2O3(0001) Surface Investigated by Scanning Tunneling Microscopy. *Surf. Sci.* **2013**, *609*, 67–72. https://doi.org/10.1016/j.susc.2012.11.005.

(5) Pollak, M.; Gautier, M.; Thromat, N.; Gota, S.; Mackrodt, W. C.; Saunders, V. R. An In-Situ Study of the Surface Phase Transitions of α-Fe2O3 by X-Ray Absorption Spectroscopy at the Oxygen K Edge. *Nucl. Instruments Methods Phys. Res. Sect. B Beam Interact. with Mater. Atoms* **1995**, *97* (1–4), 383–386. https://doi.org/10.1016/0168-583X(94)00408-0.

(6) Redondo, J.; Michalička, J.; Kraushofer, F.; Franceschi, G.; Šmid, B.; Kumar, N.; Man, O.; Blatnik, M.; Wrana, D.; Mallada, B.; Švec, M.; Parkinson, G. S.; Setvin, M.; Riva, M.; Diebold, U.; Čechal, J. Hematite α -Fe 2 O 3 (0001) in Top and Side View: Resolving Long-Standing Controversies about Its Surface Structure. *Adv. Mater. Interfaces* **2023**, *10* (32), 1–9. https://doi.org/10.1002/admi.202300602.

(7) Redondo, J.; Lazar, P.; Procházka, P.; Průša, S.; Mallada, B.; Cahlík, A.; Lachnitt, J.; Berger, J.; Šmíd, B.; Kormoš, L.; Jelínek, P.; Čechal, J.; Švec, M. Identification of Two-Dimensional FeO<inf>2</Inf> Termination of Bulk Hematite α-Fe<inf>2</Inf>O<inf>3</Inf>(0001) Surface. *J. Phys. Chem. C* **2019**, *123* (23). https://doi.org/10.1021/acs.jpcc.9b00244.

(8) Kim, C.-Y.; Escuadro, A. A.; Bedzyk, M. J.; Liu, L.; Stair, P. C. X-Ray Scattering Study of the Stoichiometric Recovery of the α-Fe2O3(0001) Surface. *Surf. Sci.* **2004**, *572* (2–3), 239–246. https://doi.org/10.1016/j.susc.2004.08.036.

(9) Tromp, R. M. Low-Energy Electron Microscopy. *IBM J. Res. Dev.* **2000**, *44*, 503–516. https://doi.org/10.1147/rd.444.0503.

(10) Bauer, E. Surface Microscopy with Low Energy Electrons: LEEM. *J. Electron Spectros. Relat. Phenomena* **2020**, *241*, 146806. https://doi.org/10.1016/j.elspec.2018.11.005.

(11) de la Figuera, J.; McCarty, K. F. Low-Energy Electron Microscopy. In *Surface Science Techniques. Springer Series in Surface Sciences, vol 51*; Bracco, G., Holst, B., Eds.; Springer Berlin Heidelberg: Berlin, Heidelberg, 2013; pp 531–561. https://doi.org/10.1007/978-3-642-34243-1_18.

(12) Nie, S.; Starodub, E.; Monti, M.; Siegel, D. A.; Vergara, L.; El Gabaly, F.; Bartelt, N. C.; de la Figuera, J.; McCarty, K. F. Insight into Magnetite's Redox Catalysis from Observing Surface Morphology during Oxidation. *J. Am. Chem. Soc.* **2013**, *135*, 10091–10098. https://doi.org/10.1021/ja402599t.

(13) McCarty, K. F.; Monti, M.; Nie, S.; Siegel, D. A.; Starodub, E.; El Gabaly, F.; McDaniel, A. H.; Shavorskiy, A.; Tyliszczak, T.; Bluhm, H.; Bartelt, N. C.; de la Figuera, J. Oxidation of Magnetite(100) to Hematite Observed by in Situ Spectroscopy and Microscopy. *J. Phys. Chem. C* **2014**, *118* (34), 19768–19777. https://doi.org/10.1021/jp5037603.

(14) Čechal, J.; Procházka, P. Low-Energy Electron Microscopy as a Tool for Analysis of Self-Assembled Molecular Layers on Surfaces. *J. Phys. Condens. Matter* **2025**, *37*, 293003. https://doi.org/10.1088/1361-648X/ade946.

(15) YANG, L. X.; MATTHEWS, E. Oxidation and Sintering of Magnetite Ore under Oxidising Conditions. *ISIJ Int.* **1997**, *37*, 854–861.

(16) Forsmo, S. P. E.; Forsmo, S.-E.; Samskog, P.-O.; Björkman, B. M. T. Mechanisms in Oxidation and Sintering of Magnetite Iron Ore Green Pellets. *Powder Technol.* **2008**, *183*, 247–259. https://doi.org/10.1016/j.powtec.2007.07.032.

(17) Yin, S.; Wirth, R.; He, H.; Ma, C.; Pan, J.; Xing, J.; Xu, J.; Fu, J.; Zhang, X.-N. Replacement of Magnetite by Hematite in Hydrothermal Systems: A Refined Redox-Independent Model. *Earth Planet. Sci. Lett.* **2022**, *577*, 117282. https://doi.org/10.1016/j.epsl.2021.117282.

(18) Zhao, J.; Brugger, J.; Pring, A. Mechanism and Kinetics of Hydrothermal Replacement of Magnetite by Hematite. *Geosci. Front.* **2019**, *10* (1), 29–41. https://doi.org/10.1016/j.gsf.2018.05.015.

(19) Tober, S.; Creutzburg, M.; Arndt, B.; Krausert, K.; Mattauch, S.; Koutsioubas, A.; Pütter, S.; Mohd, A. S.; Volgger, L.; Hutter, H.; Noei, H.; Vonk, V.; Lott, D.; Stierle, A. Observation of Iron Diffusion in the Near-Surface Region of Magnetite at 470 K. *Phys. Rev. Res.* **2020**, *2*, 023406. https://doi.org/10.1103/PhysRevResearch.2.023406.

(20) Gallagher, K. J.; Feitknecht, W.; Mannweiler, U. Mechanism of Oxidation of Magnetite to γ-Fe2O3. *Nature* **1968**, *217*, 1118–1121.

(21) Dieckmann, R.; Schmalzried, H. Defects and Cation Diffusion in Magnetite (I). *Ber. Bunsenges. Phys. Chem.* **1977**, *81*, 344.

(22) Dieckmann, R.; Schmalzried, H. Defects and Cation Diffusion in Magnetite (II). *Ber. Bunsenges. Phys. Chem.* **1977**, *81*, 414.

(23) Davis, B.; Rapp, G.; Walawender, M. Fabric and Structural Characteristics of the Martitization Process. *Am. J. Sci.* **1968**, *266*, 482–496.

(24) Cornell, R. M.; Schwertmann, U. *The Iron Oxides*; Wiley, 2003. https://doi.org/10.1002/3527602097.

(25) Freindl, K.; Wojas, J.; Kwiatek, N.; Korecki, J.; Spiridis, N. Reversible Oxidation–Reduction of Epitaxial Iron Oxide Films on Pt(111): Magnetite–Hematite Interconversion. *J. Chem. Phys.* **2020**, *152* (5). https://doi.org/10.1063/1.5136322.

(26) Dieckmann, R. Defects and Cation Diffusion in Magnetite (IV): Nonstoichiometry and Point Defect Structure of Magnetite (Fe 3 - δ O 4 ). *Berichte der Bunsengesellschaft für Phys. Chemie* **1982**, *86* (2), 112–118. https://doi.org/10.1002/bbpc.19820860205.

(27) Cabrera, N.; Mott, N. F. Theory of the Oxidation of Metals. *Reports Prog. Phys.* **1949**, *12*, 163.

(28) McIntyre, N. S.; Zetaruk, D. G. X-Ray Photoelectron Spectroscopic Studies of Iron Oxides. *Anal. Chem.* **1977**, *49* (11), 1521–1529. https://doi.org/10.1021/ac50019a016.

(29) Lad, R. J.; Henrich, V. E. Structure of α-Fe2O3 Single Crystal Surfaces Following Ar+ Ion Bombardment and Annealing in O2. *Surf. Sci.* **1988**, *193* (1–2), 81–93. https://doi.org/10.1016/0039-6028(88)90324-X.

(30) Chambers, S. A.; Kim, Y. J.; Gao, Y. Fe 2 p Core-Level Spectra for Pure, Epitaxial α -Fe2O3(0001), γ -Fe2O3(001), and Fe3O4(001). *Surf. Sci. Spectra* **1998**, *5* (3), 219–228. https://doi.org/10.1116/1.1247873.

(31) Kurtz, R. L.; Henrich, V. E. Geometric Structure of the α-Fe2O3(001) Surface: A LEED and XPS Study. *Surf. Sci.* **1983**, *129* (2–3), 345–354. https://doi.org/10.1016/0039-6028(83)90185-1.

(32) Condon, N. G.; Leibsle, F. M.; Lennie, A. R.; Murray, P. W.; Vaughan, D. J.; Thornton, G. Biphase Ordering of Iron Oxide Surfaces. *Phys. Rev. Lett.* **1995**, *75* (10), 1961–1964. https://doi.org/10.1103/PhysRevLett.75.1961.

(33) Bowker, M.; Hutchings, G.; Davies, P. R.; Edwards, D.; Davies, R.; Shaikhutdinov, S.; Freund, H.-J. Surface Structure of γ-Fe2O3(111). *Surf. Sci.* **2012**, *606*, 1594–1599. https://doi.org/10.1016/j.susc.2012.06.010.

(34) Franceschi, G.; Schmid, M.; Diebold, U.; Riva, M. Reconstruction Changes Drive Surface Diffusion and Determine the Flatness of Oxide Surfaces. *J. Vac. Sci. Technol. A* **2022**, *40* (2). https://doi.org/10.1116/6.0001704.

(35) Reuter, K.; Scheffler, M. Composition, Structure, and Stability of RuO2(110) as a Function of Oxygen Pressure. *Phys. Rev. B* **2001**, *65* (3), 035406. https://doi.org/10.1103/PhysRevB.65.035406.

(36) Chase, M. W.; Davies, C. A.; Downey, J. R.; Frurip, D. J.; McDonald, R. A.; Syverud, A. N. *NIST-JANAF Thermochemical Tables*; National Institute of Standards and Technology: Gaithersburg, MD, 1998.

TOC graphics

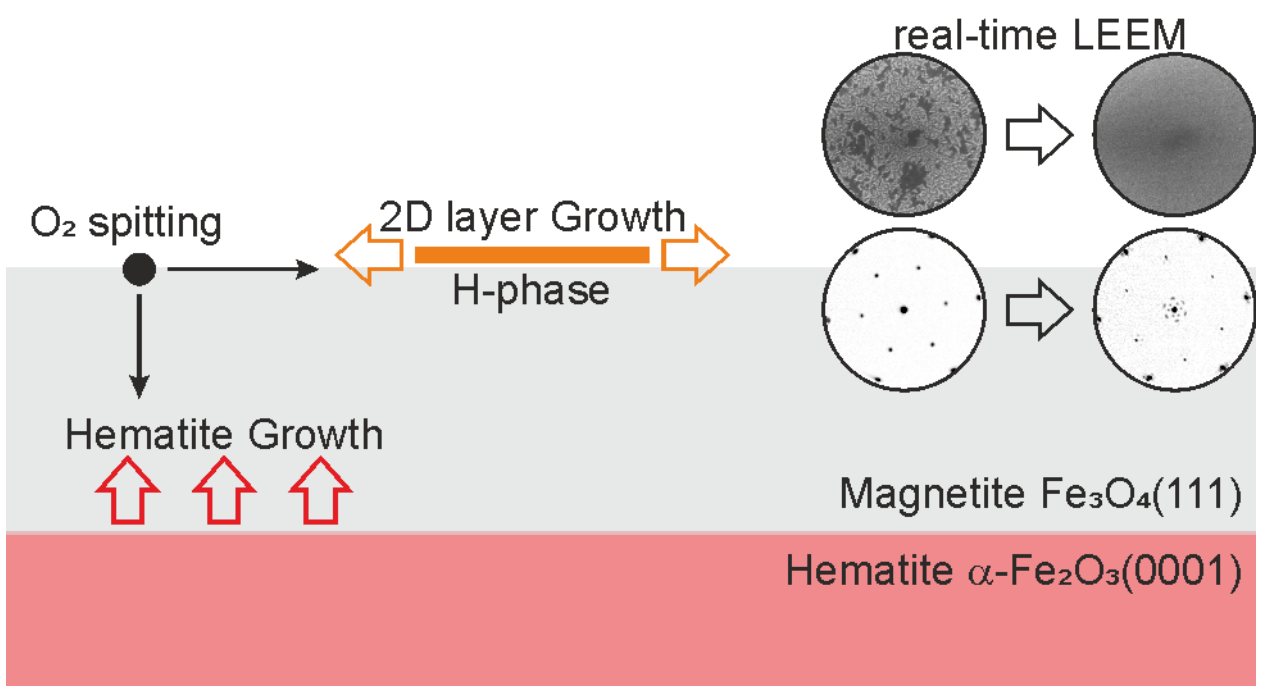

SUPPLEMENTARY INFORMATION

# Oxygen-Pressure-Limited Recovery of the Hematite $\alpha$-$Fe_2O_3$(0001) Surface from a Reduced $Fe_3O_4$(111)-Like Layer

*Nishant Kumar,[1] Matthias Blatnik,[1] Jan Čechal[1,2]**

[1] CEITEC - Central European Institute of Technology, Brno University of Technology, Purkyňova 123, 612 00 Brno, Czech Republic.

[2] Institute of Physical Engineering, Brno University of Technology, Technická 2896/2, 616 69 Brno, Czech Republic.

**Table S1**: Oxidation kinetics data for hematite ($\alpha$-$Fe_2O_3$) under different experimental conditions.

| | Temperature (°C) | Pressure (mbar) | Nucleation Time ($t_0$) [min]* | Error in $t_0$ ($\Delta t_0$) [min] | Characteristic phase growth time ($t_{50}$) [min] | Error in $t_{50}$ ($\Delta t_{50}$) [min] | Time to full oxidation ($t_{100}$) [min] |
|---|---|---|---|---|---|---|---|
| Constant Partial Oxygen Pressure | 660 | $1.8\times10^{-6}$ | 35 | 9.67 | 13.6 | 12.57 | 90 |
| | 670 | $1.8\times10^{-6}$ | 32 | 0.81 | 4.03 | 1.0 | 95 |
| | 680 | $1.8\times10^{-6}$ | 30 | 1.81 | 8.49 | 2.26 | 100 |
| | 690 | $1.8\times10^{-6}$ | 28 | 3.39 | 23.44 | 4.99 | 107 |
| | 700 | $1.8\times10^{-6}$ | 26 | 10.93 | 40.54 | 17.33 | 120 |
| Constant Temperature | 670 | $6.3\times10^{-7}$ | 165 | 18.63 | 104.65 | 26.7 | 450 |
| | 670 | $1.1\times10^{-6}$ | 65 | 16.02 | 43.18 | 23.45 | 185 |
| | 670 | $1.8\times10^{-6}$ | 32 | 0.81 | 4.03 | 1.0 | 95 |
| | 670 | $3.0\times10^{-6}$ | 21 | 19.5 | 30.21 | 26.29 | 85 |
| | 670 | $5.0\times10^{-6}$ | 18 | 1.78 | 10.14 | 2.49 | 55 |
| Constant Oxygen Chemical Potential ($\mu$=-1.867 eV) | 660 | $6.3\times10^{-7}$ | 75 | 15.49 | 104.86 | 24.18 | 355 |
| | 670 | $1.1\times10^{-6}$ | 65 | 16.02 | 43.18 | 23.45 | 185 |
| | 680 | $1.8\times10^{-6}$ | 30 | 1.81 | 8.49 | 2.26 | 100 |
| | 690 | $3.0\times10^{-6}$ | 26 | 7.33 | 15.04 | 8.48 | 95 |
| | 700 | $5.0\times10^{-6}$ | 15 | 16.09 | 16.61 | 19.27 | 60 |

*Nucleation times ($t_0$) are experimentally measured, while errors ($\Delta t_0$) are derived from the intersection of linear fit of the following linear increase with the baseline.

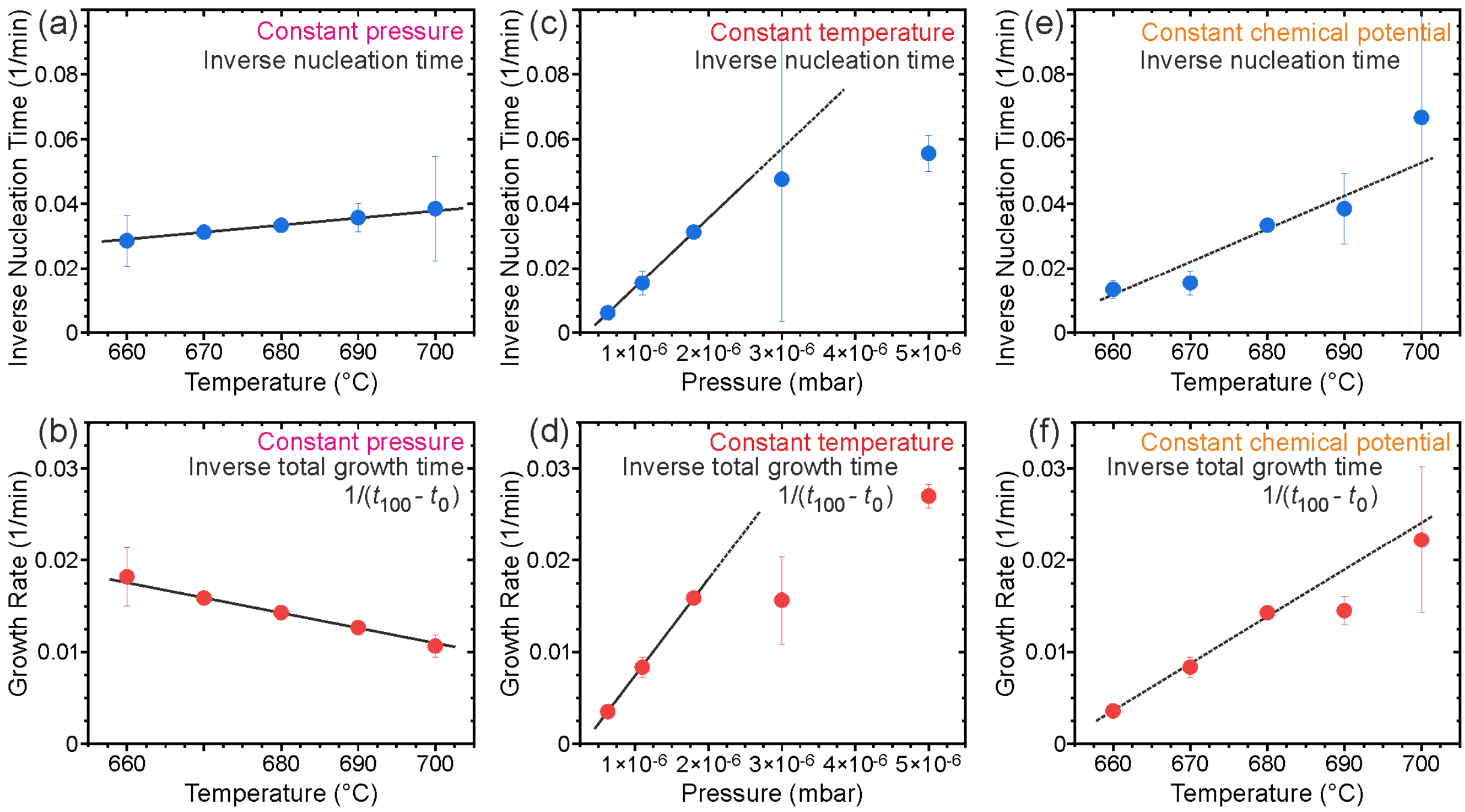

**Figure S1:** Inverse times associated with Figure 4 in the main text. (a, b) Inverse nucleation time ($t_0$) and inverse growth time $t_{100}$- $t_0$ (growth rate) as function of temperature at constant $p_{O_2} = 1.8 \times 10^{-6}$ mbar. (c, d Inverse nucleation time ($t_0$) and inverse growth time $t_{100}$- $t_0$ (growth rate) as function of $p_{O_2}$ at constant temperature (670°C). (e, f) Inverse nucleation time ($t_0$) and inverse growth time $t_{100}$- $t_0$ (growth rate) as function of temperature under constant oxygen chemical potential ($\mu_O = -1.867$ eV).

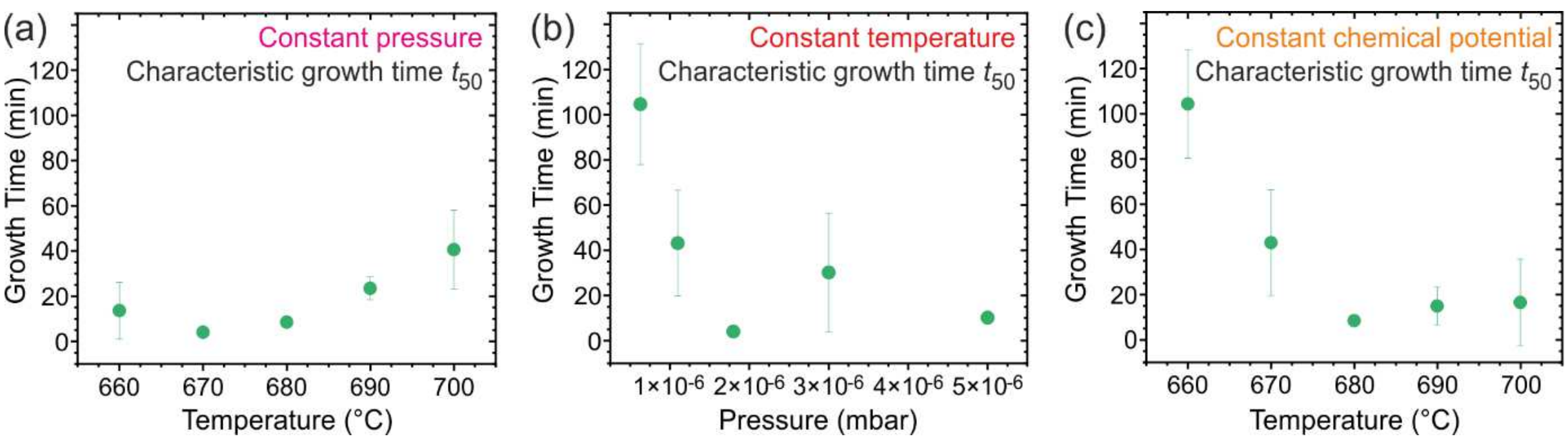

**Figure S2:** Characteristic H-phase growth time ($t_{50}$) under (a) as function of temperature at constant $p_{O_2} = 1.8 \times 10^{-6}$ mbar, (b) as function of $p_{O_2}$ at constant temperature (670°C), and (c) as function of temperature under constant oxygen chemical potential ($\mu_O = -1.867$ eV).